\documentstyle[myapj,myonecol]{article}

\def\etal{{\rm et~al.\ }}
\def\hmpc{\;h^{-1}{\rm Mpc}}
\def\invhmpc{\;h\;{\rm Mpc}^{-1}}
\def\kms{{\rm \;km\;s^{-1}}}

\input epsf

\begin{document}

\twocolumn[
\title{Suppressing Linear Power on Dwarf Galaxy Halo Scales}
\author{Martin White and Rupert A.C.~Croft}
\affil{Harvard-Smithsonian Center for Astrophysics, Cambridge, MA 02138}
\authoremail{mwhite@cfa.harvard.edu}

\begin{abstract}
\noindent
\rightskip=0pt
Recently is has been suggested that the dearth of small halos around the
Milky Way arises due to a modification of the primordial power spectrum
of fluctuations from inflation.
Such modifications would be expected to alter the formation of structure
from bottom-up to top-down on scales near where the short-scale power has
been suppressed.
Using cosmological simulations we study
 the effects of such a modification of the initial power spectrum.
While the halo multiplicity function depends primarily on the linear theory
power spectrum, most other probes of power are more sensitive to 
the non-linear power spectrum.  Collapse of large-scale 
structures as they go non-linear
regenerates a ``tail'' in the power spectrum, masking small-scale
 modifications to the primordial power spectrum except at very high-$z$.
Even the small-scale ($k > 2 \invhmpc$) 
clustering of the Ly-$\alpha$ forest is affected by this process,
so that CDM models with sufficient  power suppression to reduce the number
of $10^{10} M_{\odot}$ halos by a factor of $\sim 5$ give similar 
Ly-$\alpha$ forest power spectrum results.
We conclude  that other observations that depend more directly
on the number density
of collapsed objects, such as the number of damped Ly-$\alpha$ systems,
or the redshift of reionization may provide the  most sensitive tests
of these models. 

\end{abstract}
]

\section{Introduction} \label{sec:intro}

The field of physical cosmology has made rapid progress in the last decade,
and a ``standard model'' is already beginning to emerge.
Many of the main cosmological parameters are becoming known and there
is good reason to believe that the measurements will be significantly improved,
and the paradigm tested, in the next few years from observations of the
Cosmic Microwave Background anisotropy and upcoming surveys of large-scale
structure.
While in broad outline the paradigm appears to work well, there are some
discrepancies which indicate that revisions in our standard model may be
required.
In this paper we discuss several topics related to one of these issues:
the lack of low mass halos in our local neighborhood and in particular
consider what we might learn about the small-scale matter power spectrum.

The halo problem has been highlighted by several groups.  Analytic arguments
based on Press-Schechter~(\cite{PreSch}) theory were given by
Kauffmann, White \& Guiderdoni~(\cite{KauWhiGui}), while
Klypin et al.~(\cite{KKVP}) and Moore et al.~(\cite{MGGLQST}) used very
high resolution dark matter simulations.
A summary of the situation has been given recently by
Spergel \& Steinhardt~(\cite{SpeSte}) and
Kamionkowski \& Liddle~(\cite{KamLid}).

Within the Press-Schechter theory, and its extensions, the number density of
halos of a given mass is related to the amplitude of the linear theory power
spectrum on a scale proportional to $M^{1/3}$.
For example $10^{10}M_\odot$ halos in a model with $\Omega_{\rm m}=0.3$ probe
a linear scale of $0.3 h^{-1}$Mpc.
Numerous numerical simulations have demonstrated that halo number density
seems to be governed by the {\it linear\/} power spectrum as Press-Schechter
theory would predict.
A deficit of low mass halos thus implies either additional physics (see below)
or a deficit of linear theory power on small length scales.
This modification could come about either by variations in the primordial
power spectrum (e.g.,~from inflation) or in the cosmological processing of
this power spectrum (e.g.,~from `warm' dark matter).

Recently Kamionkowski \& Liddle~(\cite{KamLid}) pointed out that a well
studied class of inflationary models (BSI; see e.g., Starobinsky~\cite{Sta})
could give rise to a deficit of small-scale power in the primordial power
spectrum.
One way to achieve such a deficit is to introduce a change in the slope of
the inflaton potential at a scale determined by the astrophysical problem
to be solved, in our case $k\sim 5 h {\rm Mpc}^{-1}$.
Thus in such models one `naturally' achieves fewer low mass halos than in
the conventional inflationary CDM models.

The linear theory power spectrum of these BSI models is well described by
Kamionkowski \& Liddle~(\cite{KamLid}).  Compared to a scale-invariant model
there is a small rise followed by a sharp drop in power at some scale $k_0$.
Beyond $k_0$ the power spectrum oscillates with an envelope which falls more
steeply than $k^{-3}$.  At high-$k$ the spectrum recovers to the usual $k^{-3}$
slope but with much smaller amplitude.

We believe that it is of interest to constrain such modifications to the
initial power spectrum, if possible.  However, because of this sharp drop,
the model near $k_0$ more closely resembles the familiar top-down scenarios
(e.g.~HDM) than a bottom-up CDM model for some range of wavenumbers.
Thus arguments based on reasoning developed for `traditional' CDM models
should be checked against numerical simulations.
Furthermore, the number density of objects may be one of the only probes of the
linear theory power spectrum.
Several astrophysical probes of small-scale power are sensitive not to the
linear theory but to the non-linear power spectrum.
As is well known, objects collapsing under gravitational instability feed
power from large scales to small thus allowing small-scale power to be
regenerated once a mode goes non-linear
(e.g.~Little, Weinberg \& Park~\cite{LWP}; Melott \& Shandarin~\cite{MS90},
\cite{MS91}, \cite{MS93}; Bagla \& Padmanabhan~\cite{BP97}).
How much power is regenerated, crucial for determining how much there was
initially, requires numerical calculation.

We address several of these issues in the following sections.

Finally we should note that we believe that constraints on small-scale power
are of intrinsic interest in and of themselves.  We shall discuss this within
the context of the sub-halo problem described above while noting that several
other astrophysical effects may also explain the discrepancy.
The most obvious examples are: the total number of Local Group satellites
could be underestimated, feedback could be important
(Kauffmann, White \& Guiderdoni~\cite{KauWhiGui},
Bullock, Kravstov \& Weinberg~\cite{BulKraWei}),
or the satellites could fail to make stars and be dark (e.g.~HI clouds).

\section{Probes of Small-Scale Power}

Any proposal to solve the small halo number density problem by modifying
the initial power spectrum must simultaneously be able to pass other
constraints on small-scale power.  While we have a number of constraints
on the linear and evolved power spectrum on larger scales, there are very
few stringent constraints on linear scales below a Mpc.
Kamionkowski \& Liddle~(\cite{KamLid}) argue that constraints from the
abundance of damped Ly-$\alpha$ systems and the reionization epoch are
passed by low-density versions of the BSI model, partly because of the
uncertainties involved in making those predictions.
The clustering of objects at high-$z$ does not appear to be a promising probe
of the matter power spectrum on these small scales.
A priori the two most obvious probes are the object abundances which motivated
this modification of the power spectrum initially and the power spectrum of
the flux in the Ly-$\alpha$ forest.

\section{Simulations}

To address some of these issues we ran two sets of N-body simulations.
The base model in all cases was a $\Lambda$CDM model with $\Omega_{\rm m}=0.3$,
$\Omega_\Lambda=0.7$, $h=0.7$, $n=1$, COBE normalized using the method of
Bunn \& White~(\cite{BunWhi}), i.e.~with $\sigma_8=0.88$.
The transfer functions were computed using the fits of
Eisenstein \& Hu~(\cite{EisHu}) without the baryonic oscillations.
This power spectrum was optionally filtered to suppress small-scale power.
We have modeled the behavior displayed in Fig.~1 of
Kamionkowski \& Liddle~(\cite{KamLid}) with a simple analytic form:
\begin{equation}
  \Delta^2(k) \equiv {k^3P(k)\over 2\pi^2} =
  \left( \Delta^{-2}_{\rm fid} +
         (k/k_0)^{3/2} \Delta^{-2}_{\rm fid}(k_0) \right)^{-1}
\label{eqn:pkform}
\end{equation}
Here $\Delta^2_{\rm fid}$ is the fiducial power spectrum whose high-$k$
behavior we are modifying and the power-law slope of the $k/k_0$ term was
chosen to match the behavior of their Fig.~1 in the range just above $k_0$.
The model plotted in their Fig.~1 corresponds to $k_0\simeq 10h{\rm Mpc}^{-1}$,
as shown in our Fig.~\ref{fig:d2} below.

The first set of simulations used a PM code described in detail in
(Meiksin, White \& Peacock~\cite{MWP}, White~\cite{Whi}).
The simulations used $256^3$ particles and a $512^3$ force mesh in a box
$25h^{-1}$Mpc on a side evolved from $z=70$ to $z=3$.
The high mass resolution and quick execution times allowed us to explore
parameter space and address the Ly-$\alpha$ forest questions (\S\ref{sec:lya})
where very high force resolution isn't necessary.

\begin{figure*}
\begin{center}
\leavevmode
\epsfxsize=19cm \epsfbox{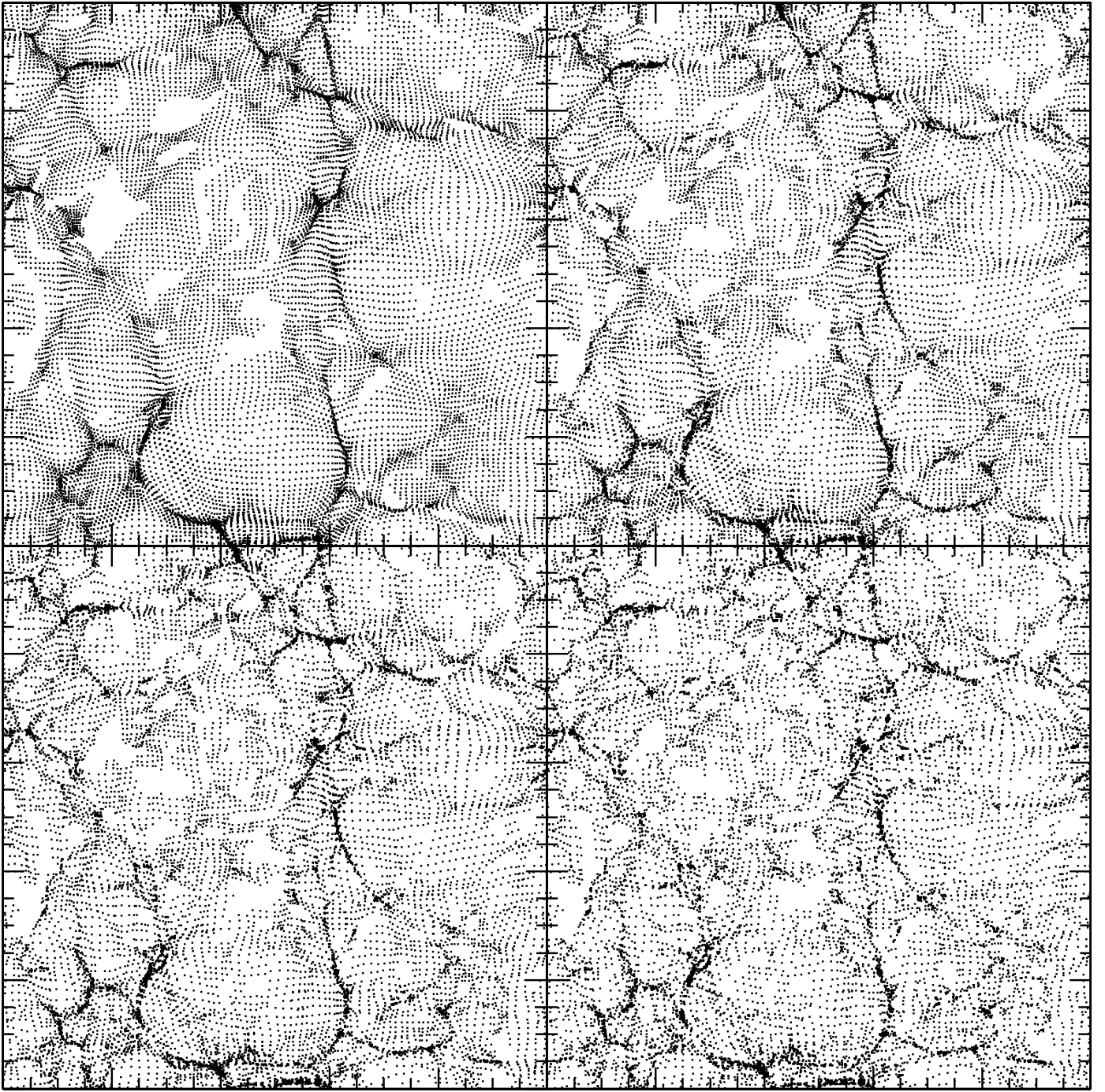}
\end{center}
\caption{Slices though the particle distribution in the {\sl TreePM\/}
simulations of our 4 models at $z=3$. From left to right, the panels show
the models with filter scale $k_0=2\invhmpc$, $k_0=5\invhmpc$ (top row),
$k_0=10\invhmpc$, and fiducial $\Lambda$CDM (bottom row).
The box side-length in each case is $25\hmpc$ and the slice thickness
$0.25\hmpc$.}
\label{fig:slice}
\end{figure*}

The second set of simulations used a new implementation of a {\sl TreePM\/}
code similar to that described in Bagla~(\cite{Bagla}).
These runs used $128^3$ particles in the same size box, evolved from
$z=60$ to $z=3$ with the time step dynamically chosen as a small fraction
of the local dynamical time.
While higher mass resolution would be preferable, this would make the
execution time prohibitive on desktop workstations with the current serial
version of the code.
A spline softened force (Monaghan \& Lattanzio~\cite{MonLat},
Hernquist \& Katz~\cite{HerKat}) with
$h=8\times 10^{-4}L_{\rm box}\simeq 20h^{-1}$kpc comoving was used
(the force was therefore exactly $1/r^2$ beyond $h$).
Very roughly this corresponds to a Plummer law smoothing $\epsilon\simeq h/3$
(e.g.~Springel \& White~\cite{SprWhi}), although a Plummer law gives 1\% force
accuracy only beyond $10\epsilon$.

We have performed numerous tests of the code, among them tests of self-similar
evolution of power-law spectra in critical density models and stable evolution
of known halo profiles.
The simulations took $\sim 200$ time steps from $z=60$ to $z=3$.
Comparison of final particle positions suggested the time step criterion was
conservative.
We have additionally compared the {\sl TreePM\/} code with a cosmological
Tree code (Springel \& White~\cite{SprWhi}) and found good agreement in the
clustering statistics for several different initial conditions, including one
of those used here (V.~Springel,, private communication).

With both the PM and {\sl TreePM\/} codes we ran 3 realizations of 4 models:
the `fiducial' $\Lambda$CDM power spectrum, and 3 filtered versions with
$k_0=10h{\rm Mpc}^{-1}$, closely approximating Fig.~1 of
Kamionkowski \& Liddle~(\cite{KamLid}),  $k_0=5h{\rm Mpc}^{-1}$
and $k_0=2h{\rm Mpc}^{-1}$ which show
a larger effect more easily resolved by these relatively small simulations.
For each of the 3 realizations the same random phases were used for all 4
power spectra to allow inter-comparison.
As additional checks on finite volume and resolution effects we also ran
simulations in boxes of side $50h^{-1}\,$Mpc and $35h^{-1}$Mpc finding
excellent agreement where the simulations overlapped.

\section{Results}

\subsection{Visual impression} \label{sec:vi}

In Fig.~\ref{fig:slice} we show slices through the particle
distributions of our 4 models. The most extreme model, with 
$k_{0}=2 \hmpc$ looks markedly different from the others,
with smooth low density regions (the simulation initial grid
is still clearly visible), and a lack of substructure in the higher density
areas. The differences between the other panels are more subtle, and in all
cases, are really only apparent on the smallest scales. 

\subsection{Power spectrum} \label{sec:pk}

\begin{figure}
\begin{center}
\leavevmode
\epsfxsize=9cm \epsfbox{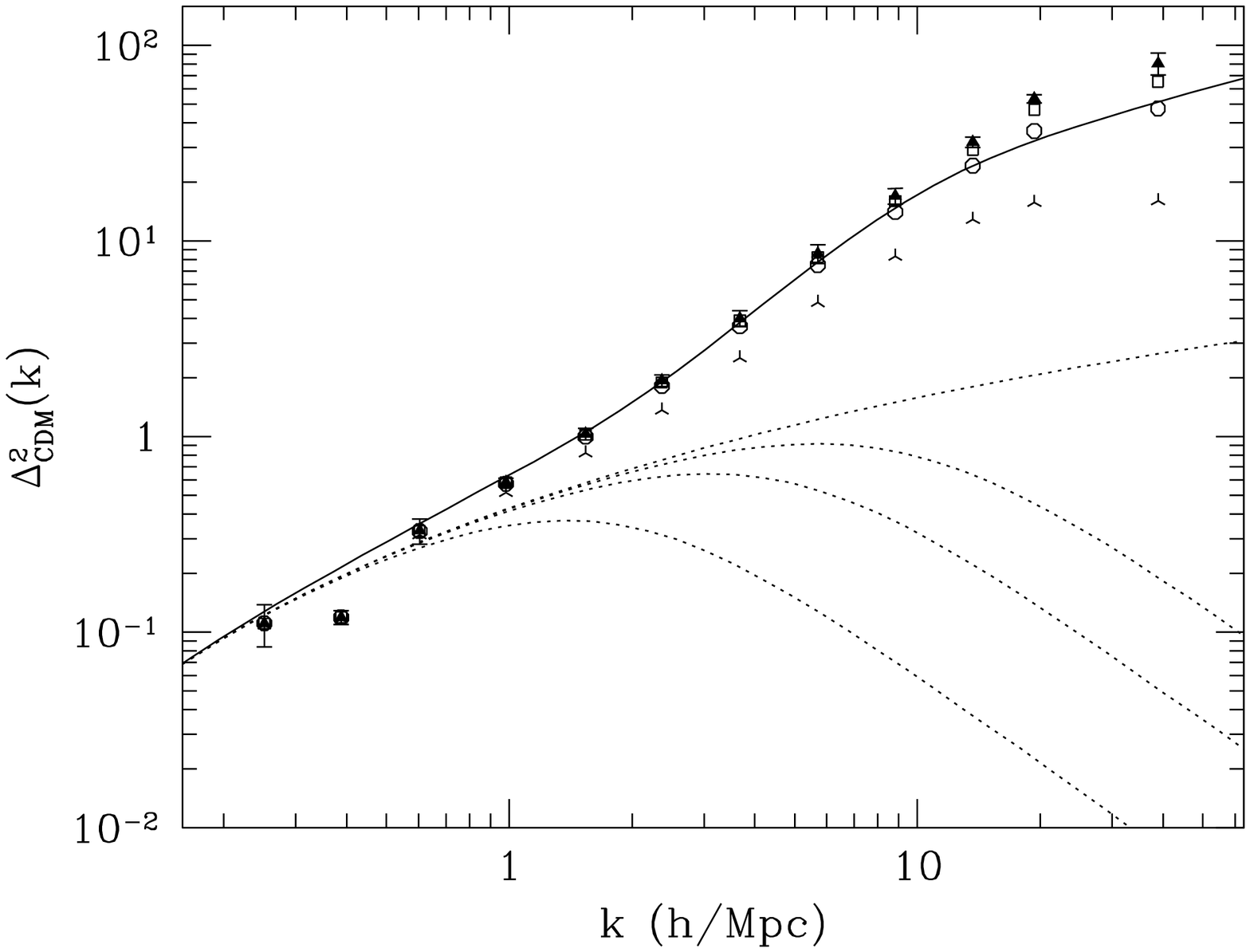}
\end{center}
\caption{The linear and non-linear power spectrum at $z=3$.
We show the mean from 3 runs of our {\sl TreePM\/} simulations for each of
4 models: our fiducial $\Lambda$CDM model (triangles), one filtered at
$k_0=10h{\rm Mpc}^{-1}$ (open squares), 
$k_0=5h{\rm Mpc}^{-1}$ (open circles), and
$k_0=2h{\rm Mpc}^{-1}$ (three pointed stars).
The solid line shows the prediction of Peacock \& Dodds~(\protect\cite{PD96})
for the fiducial model (the theory is not applicable to the filtered models)
and the dotted lines are the linear theory input spectra.
While the N-body simulations have the same random phases and so are directly
comparable, comparison with the analytic models requires error bars.
We show the $1\sigma$ error on the mean of the fiducial model computed from
our 3 realizations.  The errors on the other models are similar.}
\label{fig:d2}
\end{figure}

\begin{figure}
\begin{center}
\leavevmode
\epsfxsize=9cm \epsfbox{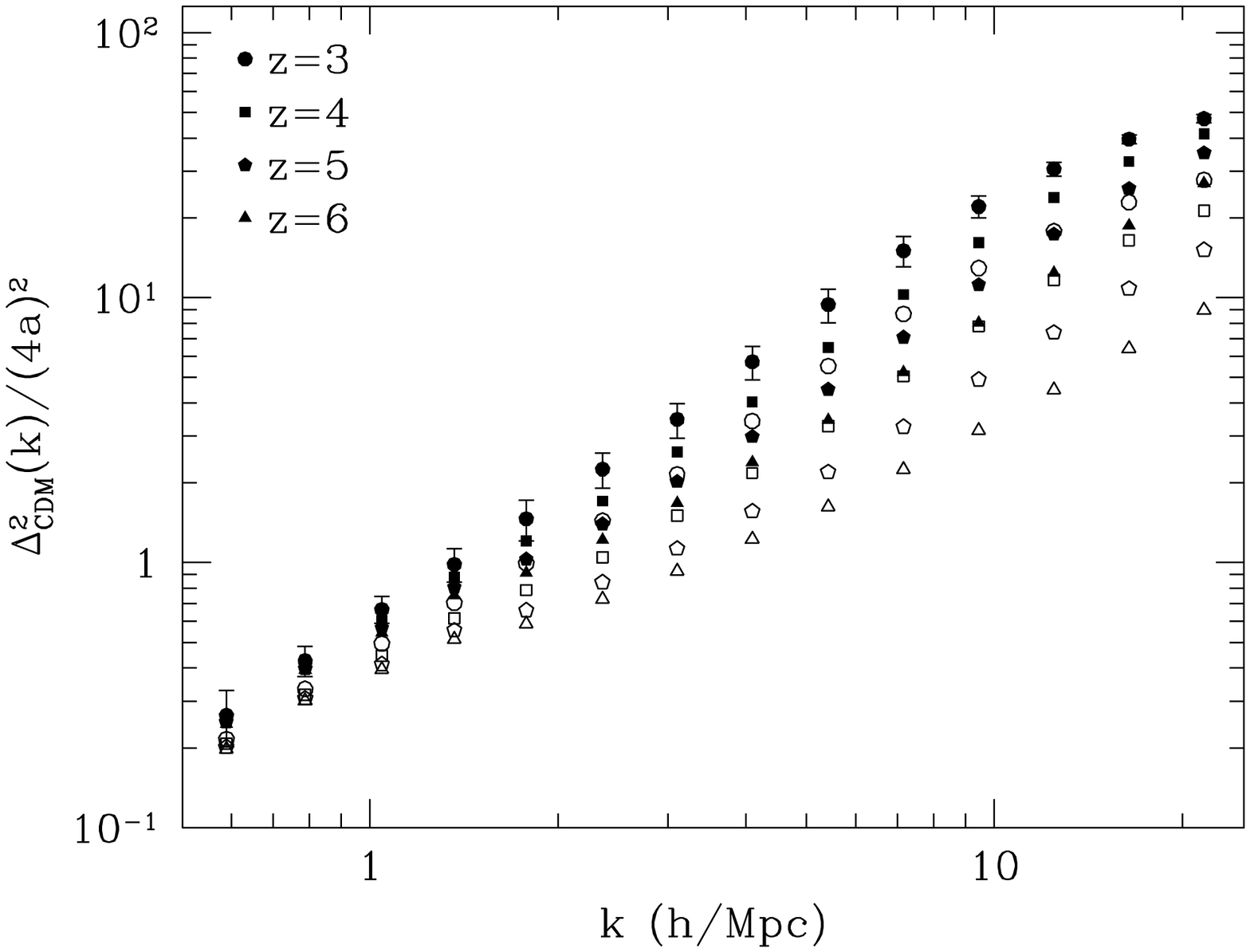}
\end{center}
\caption{The non-linear power spectra as a function of redshift for our
fiducial model (solid symbols) and one filtered model with
$k_0=5h\,{\rm Mpc}^{-1}$ (open symbols).  The symbols represent the mean
of 3 runs of our PM simulations, the error bars show the $1\sigma$ error
on the mean estimated from our 3 realizations at $z=0$.
(However recall that for each model the simulations have the same random
phases.)
The spectra are scaled by $(4a)^{-2}$ to reduce the effect of linear
evolution and highlight the non-linear growth.}
\label{fig:d2seq}
\end{figure}

Most probes of small-scale structure, other than the object abundance,
depend upon the non-linear power spectrum.
The process of gravitational collapse transfers power from large-scales
to small, and can generate a $k^{-3}$ tail in $P(k)$ if it is absent
initially.
Fitting formulae for the non-linear power spectrum such as that of
Peacock \& Dodds~(\cite{PD96}) are not applicable for spectra, such as ours,
which have regions with $n<-3$.  We use our PM and {\sl TreePM\/} simulations
to study the non-linear power spectrum.

As can be seen in Fig.~\ref{fig:d2} the scales of interest are non-linear by
$z=3$ and small-scale power removed by filtering has been regenerated by
collapse of large-scale modes.
The fiducial model has good agreement with the fitting function of
Peacock \& Dodds~(\cite{PD96}) on intermediate scales, though for
scales smaller than $k\sim 10h\,{\rm Mpc}^{-1}$ in the {\sl TreePM\/}
simulations we obtain more power than Peacock \& Dodds predict by a factor of
about 2, independent of the realization or box size.
We believe this is due to the very flat nature of the linear theory spectrum
on these scales: Jain \& Bertschinger~(\cite{JaiBer}) found a similar
discrepancy with Peacock \& Dodds for $n=-2$ spectra (see their Fig.~7).
We note that other fitting formulae have been developed which it is possible
would better reproduce the behaviour of our fiducial model
(see e.g.~Jain, Mo \& White~\cite{JMW}; Ma~\cite{CPMa98}).

To focus on the dynamics of the power regeneration we show the evolution of
the mass power spectrum for our fiducial model and one filtered model
(with $k_0=5h\,{\rm Mpc}^{-1}$) in Fig.~\ref{fig:d2seq}.
We use the average of 3 realizations of PM simulation output here since the
greater particle density allows us to probe smaller amplitude fluctuations
before shot-noise contamination becomes severe.
The PM and {\sl TreePM\/} simulations agree on the power up to
$k\sim 20h\,{\rm Mpc}^{-1}$ suggesting we resolve the relevant scales with
our PM code.

\begin{figure}
\begin{center}
\leavevmode
\epsfxsize=9cm \epsfbox{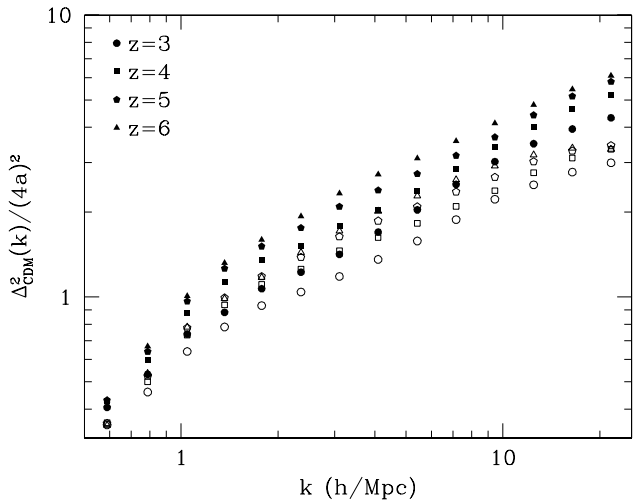}
\end{center}
\caption{The redshift space non-linear power spectra as a function of
redshift for our fiducial model (solid symbols) and one filtered model with
$k_0=5h\,{\rm Mpc}^{-1}$ (open symbols).  The symbols represent the mean
of 3 runs of our PM simulations.  The spectra are scaled by $(4a)^{-2}$ to
reduce the effect of linear evolution and highlight the non-linear growth.}
\label{fig:d2zseq}
\end{figure}

Notice that even at $z=6$, the ``peak'' in power introduced by
Eq.~\ref{eqn:pkform} has disappeared and small-scale power has been
regenerated.
The difference between the fiducial and filtered models grows progressively
smaller as the evolution proceeds.
For comparison, the generation of non-linear power has also been studied in
numerical experiments by Little, Weinberg \& Park~(\cite{LWP}), who studied
scale-invariant models,
Melott \& Shandarin~(\cite{MS90}, \cite{MS91}, \cite{MS93}) and
by Bagla \& Padmanabhan~(\cite{BP97}), amongst others.

Finally it is of interest to ask how the redshift space power spectra evolve.
Typically the redshift space spectra appear closer to the linear theory
power spectrum than the real space spectra.
In Fig.~\ref{fig:d2zseq} we show the redshift space mass power spectrum as a
function of redshift, as in Fig.~\ref{fig:d2seq} for the real-space spectra.
We can see that even the redshift space spectra have a tail of power at small
scales, induced by the non-linear clustering.

\subsection{Ly-$\alpha$ forest} \label{sec:lya}

There has been a great deal of progress in theoretical understanding of the
Ly-$\alpha$ forest recently, due in large part to hydrodynamic simulations
(Cen et al.~\cite{CMEOR}; Zhang, Anninos \& Norman~\cite{ZAN95};
Miralda-Escud\'{e} et al.~\cite{MECOR}; Hernquist et al.~\cite{HKWME};
Wadsley \& Bond~\cite{WadBon}; Zhang et al.~\cite{ZAN97};
Theuns et al.~\cite{TheLeoEfs}, \cite{TLEPT}; Dav\'{e} et al.~\cite{DHKW};
Bryan et al.~\cite{BMAN}).
In these simulations, it has been found that at high $z$ ($> 2$), 
most of the absorption in Ly-$\alpha$ forest spectra is due to a continuous, 
fluctuating photoionized medium. The physical processes governing this
absorbing gas are simple (see e.g., Bi and Davidsen~\cite{BD97},
Hui \& Gnedin~\cite{HuiGneIGM}), and as a result,
the optical depth for absorption at a particular point
can be related directly to the underlying matter density 
(Croft et al.~\cite{CWKHa}). Because of this, observations of the 
Ly-$\alpha$ forest in quasar spectra can be potentially very useful
for probing the clustering of matter (e.g.~ Gnedin~\cite{Gne98},
 Croft et al.~\cite{CWKHb},
Nusser \& Haehnelt~\cite{NusHae}).

We have generated simulated Ly-$\alpha$ forest spectra from our PM
$N$-body outputs at $z=3$ in order to test how constraining Ly-$\alpha$ 
measurements could be for the models described in this paper.
To do this, we follow a similar procedure to that outlined in
Hui \& Gnedin~(\cite{HPM}) and Croft et al.~(\cite{CWKHb}).
We bin the particle distribution onto $512^3$ density and velocity grids
using a cloud-in-cell scheme, and smooth with a Gaussian filter of width
one grid cell. 
We convert the density in each cell to an optical depth for neutral hydrogen
absorption by assuming the `Fluctuating Gunn-Peterson Approximation'' (FGPA)
(see Croft et al.~[\cite{CWKHa}, \cite{CWKHb}]; see also
Hui \& Gnedin~\cite{HPM}) and assign a temperature to the cell using a
power-law density-temperature relation.
In all our tests, we use the form $T=T_{0} \rho^{\gamma-1}$, with $\gamma=1.5$
(see Hui \& Gnedin~\cite{HuiGneIGM} for the expected
dependence of $\gamma$ on  reionization epoch).
We set the coefficient of proportionality between density and optical depth
by requiring that the mean transmitted flux, $\langle F\rangle=0.684$,
in accordance with the observations of McDonald et al.~(\cite{M99}).
We run 256 lines of sight parallel to each of the three axes through one
simulation box and create mock spectra from a convolution of the optical
depths, peculiar velocities, and thermal broadening.
The conversion to ${\rm km}\,{\rm s}^{-1}$ from $h^{-1}$Mpc at $z=3$ in this
model is a factor of 112. 

\begin{figure}
\begin{center}
\leavevmode
\epsfysize=11cm \epsfbox{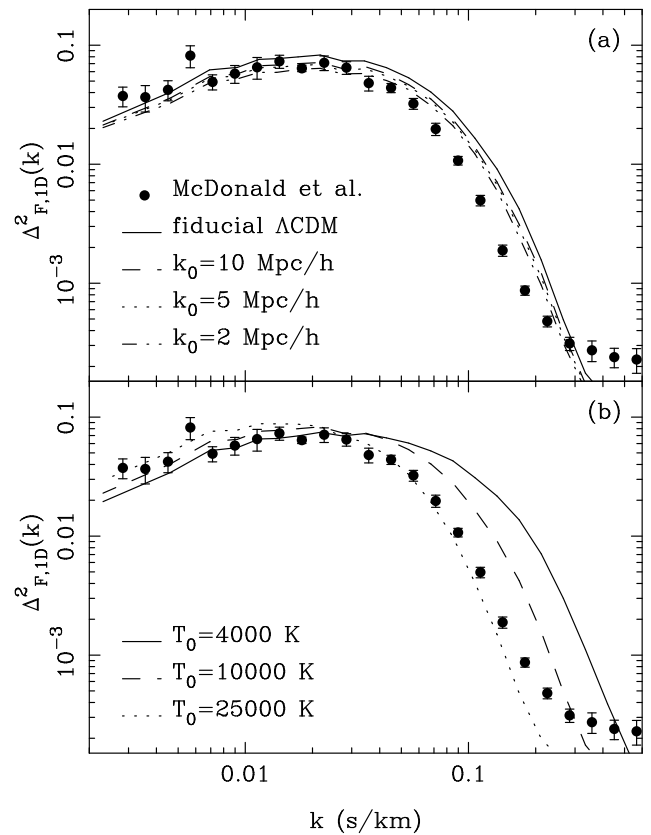}
\end{center}
\caption{(top): The 1D power spectrum of the flux measured from our simulations
of the 4 models (lines), all with $T_0=10^4$K.  The observational results, also
at $z=3$, of McDonald et al.~(1999) are shown as points.
(bottom): The 1D power spectrum of the flux for the fiducial model with 3
different values of $T_0$.}
\label{fig:fluxpk}
\end{figure}

\begin{figure}
\begin{center}
\leavevmode
\epsfxsize=8cm \epsfbox{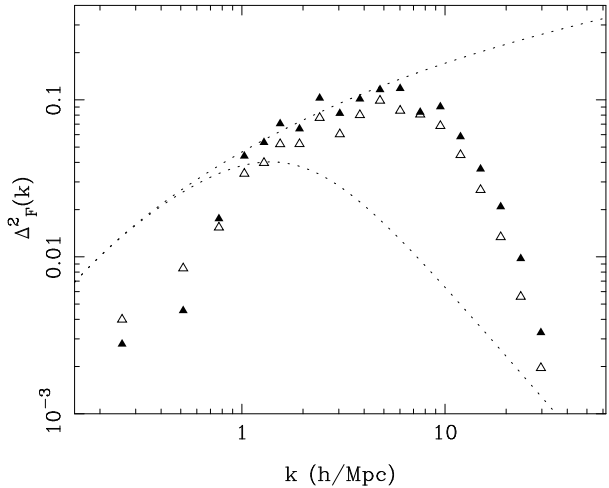}
\end{center}
\caption{The dimensionless 3D power spectrum of the flux measured from our
Ly-$\alpha$ PM simulations of 2 of our models:
fiducial $\Lambda$CDM (filled triangles), and the filtered model with
$k_0=2h{\rm Mpc}^{-1}$ (open triangles). The linear theory mass
power spectra, scaled down by an arbitrary factor to match the 
flux power spectra, are shown as dotted lines.}
\label{fig:3dfluxpk}
\end{figure}

\begin{figure}
\begin{center}
\leavevmode
\epsfxsize=8cm \epsfbox{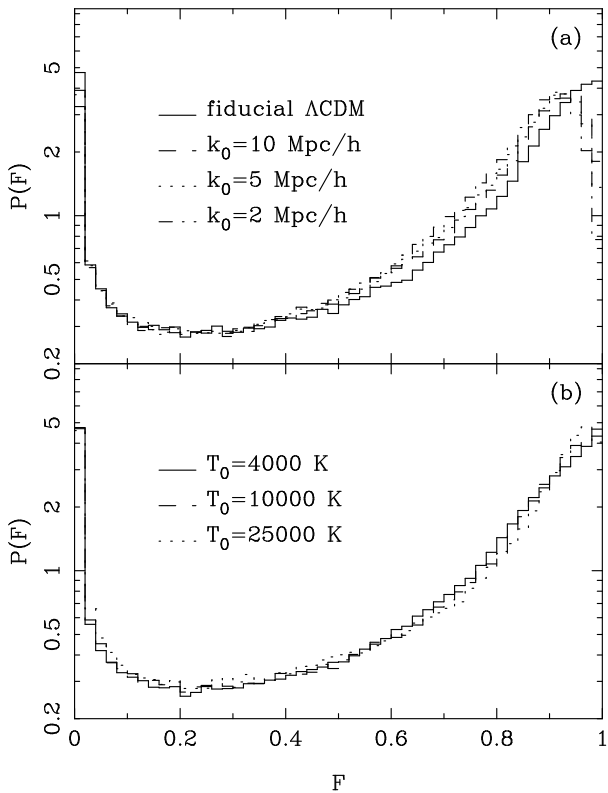}
\end{center}
\caption{Panel (a): The probability distribution function of the flux, $P(F)$,
measured from our Ly-$\alpha$ PM simulations of the 4 models, all with
$T_0=10^{4}$K.  Panel (b): $P(F)$ for the fiducial $\Lambda$CDM model, with
three different values for the temperature at the mean density.}
\label{fig:fluxpdf}
\end{figure}

On small scales, the finite pressure of the gas will in detail modify its
clustering (see e.g., Hui \& Gnedin~\cite{HPM}, Bryan et al.~\cite{BMAN},
Theuns et al.~\cite{TSH}), tending to make the gas density field smoother
than the dark matter only outputs of our simulations.
We have also implemented a 2-species version of ``Hydro-PM''
(Hui \& Gnedin~\cite{HPM}), which takes these effects into account, but find
that for our purposes here the main results are adequately reproduced by the
pure PM runs.
We expect that the details of the spectra that we create will also depend on
other assumptions about e.g., the 
reionization epoch and our simulation methodology.
To the extent that we are interested primarily in relative comparisons between
models this should not be cause for concern.

For each set of mock spectra we compute the one-dimensional flux power
spectrum, using an FFT, and show the results in Fig.~\ref{fig:fluxpk}.
In the top panel of this figure, we have set the temperature of the gas
at the mean density, $T_{0}$, to be equal to $10^4$K for all models.
The different curves show the effects of linear power suppression at different
values of $k_{0}$, while the points are the observational results of
McDonald et al.~(\cite{M99}). 
While a suppression in flux  power {\it is\/} seen in Fig.~\ref{fig:fluxpk},
it is very small.  If we vary the value of $T_0$,  the change in the thermal
broadening scale causes a more dramatic effect.
This can be seen in the lower panel of Fig.~\ref{fig:fluxpk}, where we show
results for the fiducial linear power spectrum  only.
In general, the  flux power spectrum shape on small scales will depend on the
temperature of the gas through thermal broadening and finite gas pressure, as
well as the non-linearity of matter clustering.
In the context of this $\Lambda$CDM model, it seems as though the observations
of McDonald et al.~are consistent with a fairly high mean gas temperature,
although a more detailed study involving hydrodynamic simulations is needed
to give definitive results.
What is certain from the present study is that the one-dimensional flux power
spectrum provides little constraint on our models with suppressed linear
power.
Clustering in the flux has apparently been regenerated by non-linear
gravitational evolution in a similar fashion to that seen in Fig.~\ref{fig:d2}.
There may be a good side to this, though, as insensitivity to the amount of
small-scale linear power will mean that estimates of the temperature of the
IGM made by looking  at small-scale clustering of the flux 
(as in Fig.~\ref{fig:fluxpk}) should be more robust than expected.

If we assume isotropy of clustering, the three-dimensional flux power spectrum,
$\Delta^{2}_F(k)$ can be simply recovered from the one-dimensional one
(see Croft et al.~\cite{CWKHb} for details).
It was found by Croft et al.~(\cite{CWKHb}) that on sufficiently large scales
the shape of $\Delta^{2}_F$ measured from simulated spectra matches well that
of the linear theory mass power spectrum, $\Delta^{2}(k)$.
In Fig.~\ref{fig:3dfluxpk} we test this using the model with $k_{0}=2\hmpc$,
and the fiducial model (both with $T_0=10^{4}$K).
We can see that there is not much difference between  $\Delta^{2}_F(k)$ for the
two (the same is true of the two intermediate models, which we do not plot).
The linear theory mass power spectrum (arbitrarily normalized) is shown for
comparison.  On scales approaching the box size, cosmic variance is large
enough to account for the difference between the linear curve and the points.
On smaller scales, there is still scatter, but the points taken from the
simulation with less linear power are systematically a bit lower.
We might expect the simulation points to start to trace the linear theory 
shape around the scale of non-linearity, where $\Delta^{2}(k)$ becomes
comparable to 1, which from Fig.~\ref{fig:d2} is around $k\sim1-2\invhmpc$.
If we look at Fig.~\ref{fig:3dfluxpk}, this does seem reasonable, and we do
find similar results even if we assume different gas temperatures.
On smaller scales though, $\Delta^{2}_F(k)$ has been regenerated by
non-linearity, so that the exact relationship between $\Delta^{2}_F(k)$
and the matter clustering is complex, and as in Fig.~\ref{fig:fluxpk} the
differences between models small.

Another statistic which we can check, in order to see whether suppression of
linear power has caused any changes in higher-order clustering, is the
probability distribution of the flux.
We plot this in Fig.~\ref{fig:fluxpdf}, showing the 4 models with different
linear power (and all with $T_0=10^{4}$ K) in the top panel.
There are small differences between the models, particularly at the high
flux end, where the models with more power appear to have more truly empty
regions.
These small differences are likely to remain unobserved though due to the
difficulty of accurate continuum fitting. 
This has implications for studies which use the flux PDF information to
 constrain the amount  of linear power on the Jean's scale 
(e.g., Nusser \& Haehnelt~\cite{NusHae}).
For the same reason that the flux power spectrum does not
change much on small scales (generation of power), these 
methods are likely to also be insensitive to power truncation of the
type we are considering.

The lower panel of this figure shows results for our fiducial model with
different gas temperatures.  There appears to be little difference between
the curves, although we have found that differences do appear if the spectra
are subjected to a moderate amount of smoothing
(e.g.~with a $50\kms$ Gaussian; not shown).

{}From our tests with both sets of statistics, we find that the Ly-$\alpha$
forest is not a promising discriminator between the models we are considering
here.  Two effects conspire to mask any differences in the Ly-$\alpha$
measurements on the small scales where there are large differences in the
linear power spectra.
First, the thermal broadening has the effect of smoothing the spectra 
(the thermal width of features is a few 10s of $\kms$).
Second, non-linear evolution of the density field causes power to be rapidly
transferred from large to small scales.
For these models the scale of non-linearity at $z=3$ at is about the same or
larger than the scale at which there are large differences in the linear power
spectra.
On smaller scales, the shape of the three-dimensional flux power spectrum
no longer follows that of the linear mass power spectrum.

\subsection{Halo abundance} \label{sec:halo}

\begin{figure}
\begin{center}
\leavevmode
\epsfxsize=9cm \epsfbox{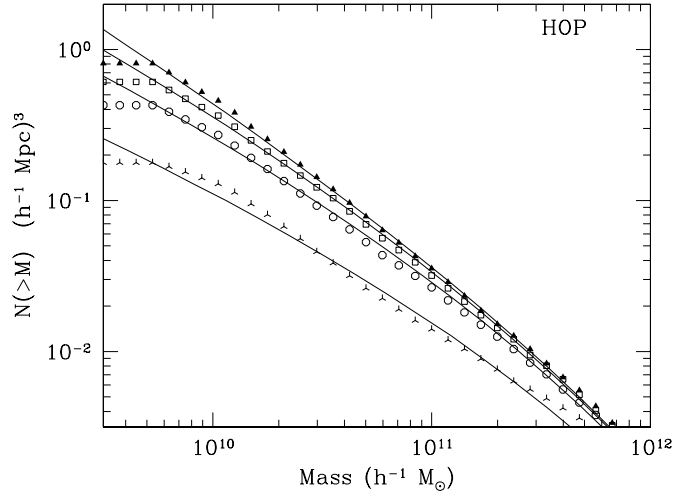}
\end{center}
\caption{ The mass functions of our simulations, using the
HOP algorithm of Eisenstein \& Hut~(\cite{HOP}) to find halos.
We show results for fiducial model (triangles) and our 3 models
with reduced small-scale power: $k_0=10h{\rm Mpc}^{-1}$ (open squares),
$k_0=5h{\rm Mpc}^{-1}$ (open circles), and
$k_0=2h{\rm Mpc}^{-1}$ (three pointed stars).
We have co-added the mass functions of the three realizations to construct
an ``average'' mass function -- the error on the mean as calculated from the 3
realizations is smaller than the size of the symbols.
The solid lines show the predictions of the Press-Schechter theory.
We have used top-hat smoothing and $\delta_c=1.5$ in the calculation of the
Press-Schechter predictions.}
\label{fig:halo}
\end{figure}

\begin{figure}
\begin{center}
\leavevmode
\epsfxsize=9cm \epsfbox{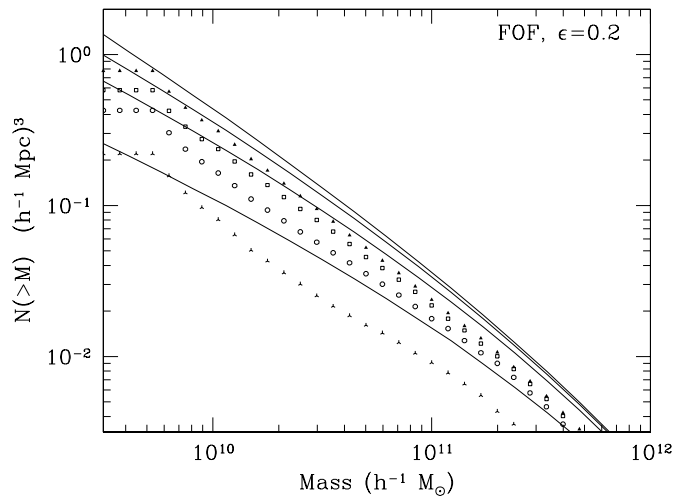}
\end{center}
\caption{The simulation mass functions, as for Fig. \ref{fig:halo}, except
that  a friends-of-friends scheme (with linking length $\epsilon=0.2$) was
used to find halos.}
\label{fig:halo_fof}
\end{figure}

Ideally we would like to evolve a large volume and study the number density
of small halos present today within a larger halo such as the Milky Way.
This is not possible with the limited dynamic range of the simulations
presented here.
While many effects {\it could\/} potentially disturb all the
small halos as they interact with
each other and a larger halo, very high resolution numerical simulations
suggest that this may not be the case in practice.
Moore et al.~(\cite{MGGLQST}) find that a large number
of small halos are not disrupted, so that the
number remaining will still be a substantial fraction of
 the number that existed in the proto-galaxy.
In this paper, we focus on the number density of halos in our simulations at
$z=3$ where our $25h^{-1}$Mpc box is just about to go non-linear.
We assume that these small halos would become incorporated into a
larger halo at later times by the usual evolution of clustering,
and that the fraction that survive disruption can be predicted
by the referring to the detailed calculations of Moore et al.~(\cite{MGGLQST}).
Here we are only interested in deficit in the number of small 
halos in our suppressed models relative to the fiducial model. 
This relative fraction should be similar at the redshift 
of our simulation box to what it would be at $z=0$, although the
absolute number of halos could only be quanitified using 
simulations like those of Moore et al. 
Similar assumptions to ours 
were also used by Kamionkowski \& Liddle~(\cite{KamLid}).

We show in Fig.~\ref{fig:halo} the Press-Schechter predictions for our
fiducial and filtered models and the numerically determined mass functions.
As there is no perfect algorithmic definition of a ``group'' of points, the
mass function is sensitive to a small degree to the halo finding algorithm.
We have used both the Friends-of-Friends (FOF; Davis et al.~\cite{DEFW})
algorithm, with linking length 0.2, and the HOP (Eisenstein \& Hut~\cite{HOP})
halo finding algorithm to construct these mass functions.
We find that the mass function differed slightly if we changed the parameters
in the algorithms or the algorithm used, and show results for both of these
schemes.  However, these differences should not affect our main conclusions,
as we are interested in comparing models with different amounts of small-scale
power to each other.

The N-body mass functions and the Press-Schechter predictions are shown in
Figs.~\ref{fig:halo} and ~\ref{fig:halo_fof} for the HOP and FOF algorithms
respectively.
The ``shelf'' at the low mass end of the N-body mass functions arises because
of the minimum number of particles allowed to form a group.  There are no
very low mass halos in the simulation.
If we use $\delta_c=1.69$ with a top-hat window in the Press-Schechter
predictions we find that the mass functions have too few large-mass halos
compared to the HOP N-body results for both $z=4$ and 3. 
The two can be brought into better agreement if we decrease $\delta_c$ to 1.5
as we have done.
With this modification, the Press-Schechter predictions overestimate the FOF
results by a factor of up to 2.
To check for simulation artifacts we also ran several larger boxes.
We find that the mass functions from a sequence of larger boxes
(up to $50h^{-1}$Mpc) with different random phases match smoothly and stably
onto the mass function of these simulations, suggesting that there are no
finite volume or sample variance effects operating.
As a final check, a completely separate analysis chain using a different N-body
code (a cosmological Tree code) obtains the same mass function at $z=3$ for one
of our runs (V.~Springel; private communication).

Figs.~\ref{fig:halo} and ~\ref{fig:halo_fof} suggest that the number of halos
is indeed governed by the linear theory power spectrum.  The amount of
suppression relative to the fiducial model is robust to the parameters of our
group finding algorithm or the algorithm used.  The absolute number of halos
can in principle be predicted from statistics of the initial density field,
although there are uncertainties  related to the definition of halos in the
simulations and parameters in Press-Schechter theory.

We find that in order to reduce the number of small halos by a large factor
(for example Kamionkowski \& Liddle~(\cite{KamLid}) recommend about an order
of magnitude), we require a fairly severe filtering of the fiducial model,
using a filter with $k_0=2\invhmpc$.

Finally we remark that this set of simulations does not have enough mass
resolution to probe the structure of the halos we find.  However simulations
by Moore et al.~(\cite{MQGSL}) suggest that the halo structure will not be
sensitive to the filtering of the initial power spectrum.  This lends some
support to our assumption that the amount of disruption of the small halos
when they become incorporated into a larger halo does not depend on the
alterations we have made to the initial power spectrum.

\section{Conclusions}

While the essential picture of hierarchical formation of large-scale
structure in a universe containing primarily cold dark matter appears to work
well, some puzzles remain.  One of these is the paucity of dwarf galaxies in
the local neighborhood.
One resolution of this ``lack-of-small-halos problem'' is a modification of
the initial power spectrum, reducing the amount of small-scale power.
There exist inflationary models which can accomplish this, though the scale of
the modification must be put in by hand.
Other approaches, such as assuming that the universe is dominated by
Warm Dark Matter (WDM), will have a similar effect (and both approaches may
solve other problems: see e.g.~Sommer-Larsen \& Dolgov~\cite{SLD99}).
In models with reduced small-scale power structure forms in a top-down manner
over a range of scales near the break, so ansatz\"{e} developed for the
``traditional'' bottom-up scenario should be treated with caution.
In this work we have used numerical simulations to address the question of
how one could constrain such a modification of the initial power spectrum.
We note that we have dealt in detail only with a model with 
suppressed initial power. In a WDM model the deficit of
power arises from the dark matter velocity dispersion, and so such as model
may behave slightly differently, at least on the smallest scales.

We find that the halo mass function depends primarily on the linear theory
power spectrum, so a suppression of small-scale power does reduce the number
of low mass halos.  While the Press-Schechter theory predicts qualitatively
the right behavior, its free parameter ($\delta_c$) must be adjusted to fit
the N-body results.  To reduce the number of $10^{10}M_\odot$ halos by a factor
of $>5$ compared to our fiducial model requires a fairly extreme filtering of
the primordial power spectrum, and the structure that forms in such a model
appears qualitatively different to the fiducial $\Lambda$CDM model
(Fig.~\ref{fig:slice}).

Collapse of large-scale structures as they go non-linear regenerates a ``tail''
in $P(k)$ if it is suppressed in the initial conditions (and this holds in
redshift as well as real space).  Thus probes which measure primarily the
evolved power spectrum are less sensitive to reduced small-scale power than one
might think.  We particularly examine measurements of clustering from the
Ly-$\alpha$ forest flux.  On the scales which govern the number of small halos,
choosing a different gas temperature affects Ly-$\alpha$ clustering much more
strongly than suppressing the linear power spectrum. The matter power spectrum
measurement made from the low resolution Ly-$\alpha$ forest spectra by
Croft \etal~(\cite{CWPHK}) probes scales just above this, which are still
linear, and offers essentially little constraint on these models.
Any extension of these simple Ly-$\alpha$ forest measurements to smaller
scales must necessarily have less general conclusions drawn from them.

Given that the number density of collapsed objects seems to be the most
sensitive probe of this small-scale modification of the power spectrum,
other observations which depend on this should be used to make consistency
checks.  At the moment, the obvious choices, such as the number density of
damped Ly-$\alpha$ systems, or the redshift of reionization induced by the
formation of the first stars and quasars, are difficult to predict accurately
{}from  theory. Their potentially strong discriminatory power will make them
useful eventually though, as we learn whether more of Cosmology's puzzles can
be resolved by an absence of small-scale power.

\bigskip
\acknowledgments  
M.W.~would like to acknowledge useful conversations with Jasjeet Bagla,
Chip Coldwell, Lars Hernquist and Volker Springel on the development of
the {\sl TreePM\/} code.  We thank Marc Kamionkowski and David Weinberg
for useful comments on an earlier draft.
M.W.~was supported by NSF-9802362 and R.A.C.C.~by 
NASA Astrophysical Theory Grant NAG5-3820.
Parts of this work were done on the Origin2000 system at the National
Center for Supercomputing Applications, University of Illinois,
Urbana-Champaign.

\end{document}